\documentclass[aps,prl,twocolumn,epsfig,groupedaddress]{revtex4}

\usepackage{epsfig}
\begin{document}

\newcommand{\qs}{Q_{\rm sat}}
\newcommand{\qsa}{Q_{\rm sat, A}}

\title{Relating high-energy lepton-hadron, proton-nucleus and 
nucleus-nucleus collisions through geometric scaling}
\author{N\'estor Armesto, Carlos A. Salgado and Urs Achim Wiedemann}

\affiliation{Department of Physics,
CERN, Theory Division, CH-1211 Gen\`eve 23, Switzerland}
\date{\today}

\begin{abstract}
A characteristic feature of small-$x$ lepton-proton
data from HERA is geometric scaling $-$ the fact that 
over a wide range of $x$ and $Q^2$ all data can be described by a 
single variable $Q^2/\qs^2(x)$, with all $x$-dependence encoded in 
the so-called saturation momentum $\qs(x)$. Here, we observe that 
the same scaling ansatz accounts for nuclear photoabsorption cross 
sections and favors the nuclear dependence $\qsa^2\propto A^{\alpha}\qs^2$, 
$\alpha \simeq 4/9$. We then make the empirical finding that 
the same $A$-dependence accounts for the centrality evolution 
of the multiplicities measured in Au+Au collisions at RHIC. 
It also allows to parametrize the high-$p_t$ particle suppression 
in d+Au collisions at forward rapidities. If these geometric scaling 
properties have a common dynamical origin, then this $A$-dependence of 
$\qsa^2$ should emerge as a consequence of the underlying 
dynamical model.
\end{abstract}
\maketitle


\noindent 1. All data for the photoabsorption cross section
$\sigma^{\gamma^* p}(x, Q^2)$ in lepton-proton scattering 
with $x\leq 0.01$ have been found~\cite{Stasto:2000er} to lie 
on a single curve when plotted against the variable $Q^2/\qs^2$, 
with $\qs^2\sim x^{-\lambda}$ and $\lambda\simeq 0.3$. To further
explore this empirical property of {\it geometric scaling}, we 
study here how experimental data on lepton-nucleus collisions constrain 
the geometric information entering $\qs^2$. We also ask to
what extent the geometric scaling ansatz can account for 
characteristic features of particle production in other
nuclear collision systems.

Geometric scaling is usually motivated in the QCD dipole 
model~\cite{dipole} where the total $\gamma^*h$ cross section
reads
\begin{eqnarray}
  \sigma_{T,L}^{\gamma^* h}(x,Q^2) = \int d^2{\bf r} \hspace{-0.1cm}
  \int_0^1
  \hspace{-0.2cm} dz
  \vert \Psi_{T,L}^{\gamma^*}(Q^2,{\bf r},z)\vert^2\,
  \sigma_{\rm dip}^h({\bf r},x)\, .
  \label{eq1}
\end{eqnarray}
Here $\Psi_{T,L}$ are the perturbatively computed transverse and 
longitudinal wave functions for the splitting of $\gamma^*$ into 
a $q\bar{q}$ dipole of transverse size ${\bf r}$ with light-cone 
fractions $z$ and $(1-z)$ carried by the quark and antiquark respectively.
Both, for a proton [$h=p$] and for a nucleus [$h=A$], 
$\sigma_{\rm dip}^h({\bf r},x)$ can be written as an integral of 
the dipole scattering amplitude $N_h$ over the impact parameter 
${\bf b}$,
\begin{eqnarray}
  \sigma_{\rm dip}^h({\bf r},x) = 2 \int d^2{\bf b}\,
                                     N_h({\bf r},x;{\bf b})\, .
  \label{eq2}
\end{eqnarray}
In this setting, geometric scaling corresponds to the
condition $N_h({\bf r},x;{\bf b})\equiv N(r\qs(x,{\bf b}))$. 
This can be seen by rescaling the impact parameter in (\ref{eq1})
in terms of the radius $R_h$ of the hadronic target, 
${\bf \bar b}={\bf b}/\sqrt{\pi R_h^2}$,
\begin{eqnarray}
  \sigma_{T,L}^{\gamma^* h}(x,Q^2) &=&  \pi R_h^2
 \int d^2{\bf r} \hspace{-0.1cm}
  \int_0^1
  \hspace{-0.2cm} dz
  \vert \Psi_{T,L}^{\gamma^*}(Q^2,{\bf r},z)\vert^2 \nonumber \\
  &\times& 2 \int d^2{\bf \bar b}\, N_h(r\qs(x,{\bf \bar b}))\, .
\label{eqnuclsc}
\end{eqnarray}
Since $\vert \Psi_{T,L}^{\gamma^*}(Q^2,{\bf r},z)\vert^2$ is
proportional to $Q^2$ times a function of ${\bf r}^2 Q^2$, 
Eq. (\ref{eqnuclsc}) depends solely on $\tau = Q^2 / \qs^2(x,{\bf \bar b})$. 
In the case of $\gamma^*$--$A$ interactions, 
geometric scaling is the property that the $A$-dependence of the
ratio $\sigma_{T,L}^{\gamma^* h}/\pi R_h^2$ can be absorbed in the
$A$-dependence of $\qsa(x,{\bf \bar b})$,
\begin{equation}
  \frac{\sigma^{\gamma^*A}(\tau_A)}{\pi R_A^2}=
  \frac{\sigma^{\gamma^*p}(\tau_p=\tau_A)}{\pi R_p^2}\, .
  \label{eqnormal}
\end{equation}
For this $A$-dependence, we make the ansatz that the saturation
scale in the nucleus grows with the quotient of the transverse
parton densities to the power $1/\delta$,
\begin{equation}
   Q_{\rm sat,A}^2=Q_{\rm sat,p}^2\left(\frac{A \pi R_p^2}
   { \pi R_A^2}\right)^\frac{1}{\delta} \hspace{-0.1cm}
   \Rightarrow 
   \tau_A=\tau_h\left(\frac{ \pi R_A^2}{A 
                       \pi R_h^2}\right)^\frac{1}{\delta} ,
   \label{eqtaua}
\end{equation}
where the nuclear radius is given by the usual parametrization
$R_A=(1.12 A^{1/3}-0.86 A^{-1/3})$ fm. We treat $\delta$ and 
$\pi R_p^2$ as free parameters to be fixed by data. 
 
\begin{figure}[hbt]
\epsfxsize=7.cm
\centerline{\epsfbox{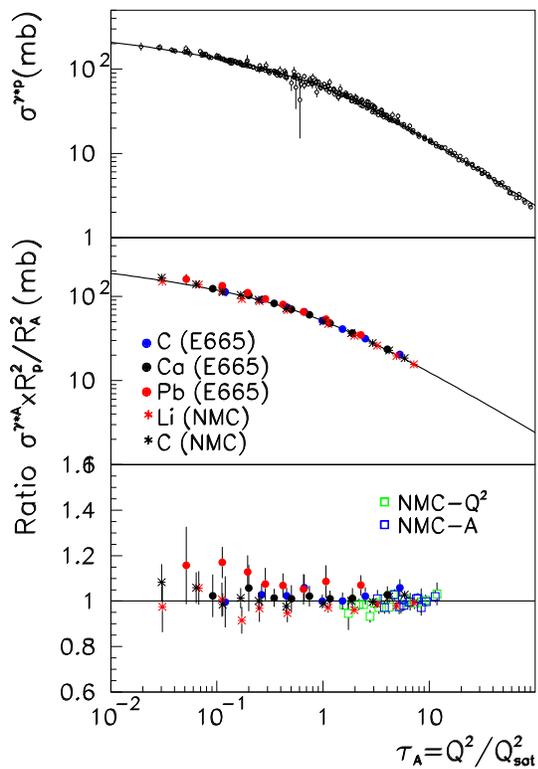}}
\caption{Geometric scaling for $\gamma^*p$ (upper panel, data
from~\cite{proton}),
$\gamma^*A$ (middle panel, data from \cite{Adams:1995is,Arneodo:1995cs}) 
and the ratio of data for $\gamma^*A$ divided by the scaling 
curve (\ref{eqscalf}) (lower panel). Also shown in the lower panel 
are the data from \cite{Arneodo:1996rv} for ratios over $C$.}
\label{figprot}
\end{figure}

\noindent 2. In Fig.~\ref{figprot} we plot the experimental $\gamma^*p$ 
data~\cite{proton} with $x\leq 0.01$ as a function of 
$\tau= Q^2 / \qs^2$. For $\qs^2$, we use in this plot
the  Golec-Biernat and W\"usthoff (GBW) 
parametrization~\cite{Golec-Biernat:1998js} 
with $R_0^2=1/\qs^2=(\bar x/x_0)^\lambda$ in GeV$^{-2}$,
$x_0= 3.04\cdot 10^{-4}$ and $\lambda=0.288$.  
To safely extend to low virtuality, the $x$-dependence of the
GBW parametrization is modified by a mass term 
$\bar x=x\left(\frac{Q^2+4m_f^2}{Q^2}\right)$ with $m_f=0.14$ GeV. 
The data~\cite{proton} are seen to be parametrized well by 
the scaling curve
\begin{eqnarray}
  \sigma^{\gamma^* p}(x,Q^2) \equiv
  \Phi(\tau) = 
\bar\sigma_0
  \left[ \gamma_E + \Gamma\left(0,\xi\right) +
         \ln\xi \right]\, ,
       \label{eqscalf}
\end{eqnarray}
where $\gamma_E$ is the Euler constant, $\Gamma\left(0,\xi\right)$
the incomplete $\Gamma$ function and $\xi=a/\tau^b$,
with $a=1.868$ and $b=0.746$. The normalization is fixed by
$\bar\sigma_0=40.56$ mb. The functional shape of (\ref{eqscalf})
can be motivated by making simplifying assumptions in
evaluating (\ref{eq1}). For our purpose, however, Eq.~(\ref{eqscalf}) 
is just a convenient ansatz for the scaling function $\Phi(\tau)$. 

To determine $\qsa^2$, we compare the functional shape of 
(\ref{eqscalf}) to the available experimental data for $\gamma^*A$ 
collisions with $x\leq 0.0175$ 
\cite{Adams:1995is,Arneodo:1995cs,Arneodo:1996rv}, using 
$\xi = a/\tau_A^b$.
The parameters $\delta$ and $\pi R_p^2$ in 
(\ref{eqnormal}) -- (\ref{eqscalf}) are fitted by $\chi^2$ 
minimization adding the statistical 
and systematic errors in quadrature. The data sets \cite{Adams:1995is}, 
\cite{Arneodo:1995cs} and 
\cite{Arneodo:1996rv} have additional normalization errors of 0.4\%, 0.2\% and 
0.15\%; the quality of the fit improves by multiplying the data
by the factors 1.004, 1.002 and 0.9985 respectively. We
obtain $\delta=0.79\pm0.02$ and $\pi R_p^2=1.55 \pm 0.02$
fm$^2$ for a $\chi^2/{\rm dof} = 0.95$ -- see Fig.~\ref{figprot} for 
comparison. If the normalizations are all set to 1,
we obtain an almost identical fit with $\delta=0.80\pm 0.02$
and $\pi R_p^2=1.57 \pm 0.02$ fm$^2$ for a $\chi^2/{\rm dof} = 1.02$. 
If we impose $\delta=1$ in the fit, which corresponds to $Q_{\rm
sat}^2\propto A^{1/3}$ for large nuclei, a much worse value of
$\chi^2/{\rm dof} =2.35$ is obtained. We conclude that the small-$x$
experimental data on $\gamma^*A$ collisions favor an increase of 
$\qsa^2$ faster than $A^{1/3}$. The numerical coincidence 
$b\simeq \delta$ is consistent with the absence of shadowing
in nuclear parton distributions at $Q^2\gg \qsa^2$.

\noindent 3.
Can geometric scaling, and in particular the $A$-dependence and energy 
dependence of $\qsa(x)$, account for the $p_t$-integrated multiplicity 
in symmetric nucleus-nucleus collisions at mid-rapidity? To address 
this question, we turn now to the heuristic ansatz
\begin{equation}
  \frac{dN^{AA}}{dy}\Bigg\vert_{y\sim 0}\propto 
  Q_{\rm sat,A}^2\, \pi R^2_A\, ,
\label{eqmprop}
\end{equation}
which arises in several models of 
hadroproduction~\cite{Gribov:tu,facto,Eskola:1999fc,Kovchegov:2000hz}. These
models relate the parton distribution measured in $\sigma^{\gamma^* A}$
to the hadroproduction measured in nucleus-nucleus collisions.
For example, the factorized formula~\cite{Gribov:tu} calculates gluon 
production by convoluting $A$-dependent gluon distribution functions 
\begin{equation}
   \frac{dN^{AB}_g}{dyd^2p_td^2b}\propto \frac{\alpha_S}{p_t^2}
   \int d^2k\  \phi_A(y,k^2,b)\,\phi_B\left(y,(k-p_t)^2,b\right)\, ,
   \label{eqfact}
\end{equation}
where $\phi_h(y,k,b)=\int d^2 r\,\exp\{i{\bf r\cdot k}\}\,
N_h({\bf r},x;{\bf b})/(2\pi r^2)$~\cite{facto}.
For geometric scaling, $\phi_A(y,k^2,b)$$\equiv$
$\phi(k^2/\qsa^2(y,b))$, we find the dependence of
Eq.~(\ref{eqmprop}),
\begin{eqnarray}
  &&\frac{dN^{AA}_g}{dy}\Bigg\vert_{y\sim 0}\propto 
  \int \frac{d^2p_t}{p_t^2} dk^2 d^2b\ 
  \phi\left(\frac{k^2}{\qsa^2}\right)\,
  \phi\left(\frac{(k-p_t)^2}{\qsa^2}\right)\nonumber \\
  &&=\qsa^2 \pi R_A^2\int \frac{d^2s}{s^2} d^2\tau d^2\bar b\ 
  \phi(\tau^2)\,\phi\left( (\tau-s)^2\right).
  \label{eqsym}
\end{eqnarray}
Also without invoking factorization in (\ref{eqfact}), 
any integrand with $(k/\qsa)$-scaling leads
to Eq.~(\ref{eqmprop}), see~\cite{Eskola:1999fc,Kovchegov:2000hz}.
In all these models, a one-to-one correspondence between parton and
hadron yields is assumed. 

\begin{figure}
\epsfxsize=7cm
\centerline{\epsfbox{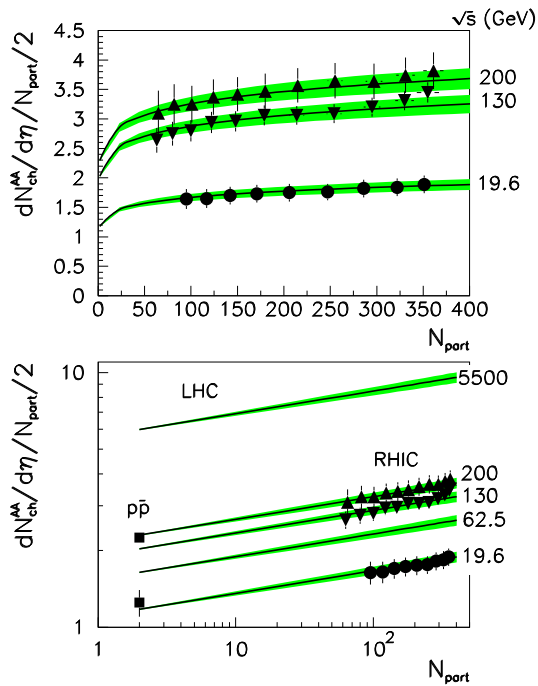}}
\caption{Energy and centrality dependence of the multiplicity of charged 
particles in AuAu collisions (\ref{eqmult}) compared to PHOBOS data
\cite{Back:2002uc}.
Also shown in the lower panel are the ${\bar p}p$ data 
\cite{protmult} and results for $\sqrt{s}=$ 62.5 and 5500 GeV/A.}
\label{figmult}
\end{figure}

To write (\ref{eqmprop}) in measurable
quantities, we express the energy dependence of the saturation 
scale in terms of the GBW parameter $\lambda=0.288$, and we translate
its $A$-dependence to an $N_{\rm part}$ dependence fixed by our 
fit parameter $\delta=0.79\pm 0.02$,
\begin{equation}
\frac{1}{N_{\rm part}}
\frac{dN^{AA}}{d\eta}\Bigg\vert_{\eta\sim 0}=N_0\sqrt{s}^\lambda 
N_{\rm part}^{\frac{1-\delta}{3\delta}}\, .
\label{eqmult}
\end{equation}
The overall normalization, independent of the energy and the centrality 
of the collision, is fixed to $N_0=0.47$. As seen in Fig.~\ref{figmult}, 
this reproduces without further adjustment experimental data 
from the PHOBOS Collaboration~\cite{Back:2002uc} on charged multiplicities 
in Au+Au collisions at $\sqrt{s}= 19.6$, 130 and 200 GeV/A. 
Even the ${\bar p}p$ data (\cite{protmult}, as quoted in \cite{Back:2002uc})
at $\sqrt{s}=$ 19.6 and 200 GeV are accounted for by Eq.~(\ref{eqmult}). 
In the same figure, we show the result of (\ref{eqmult}) 
for intermediate RHIC energy ($\sqrt{s}= 62.5$), for LHC energy 
($5500$ GeV/A) and for smaller colliding nuclei. Eq.~(\ref{eqmult}) 
implies that the energy and the centrality dependence of the 
multiplicity factorize, in agreement with the results by 
PHOBOS~\cite{Back:2002uc}.
%

\begin{figure}
\epsfxsize=7cm
\centerline{\epsfbox{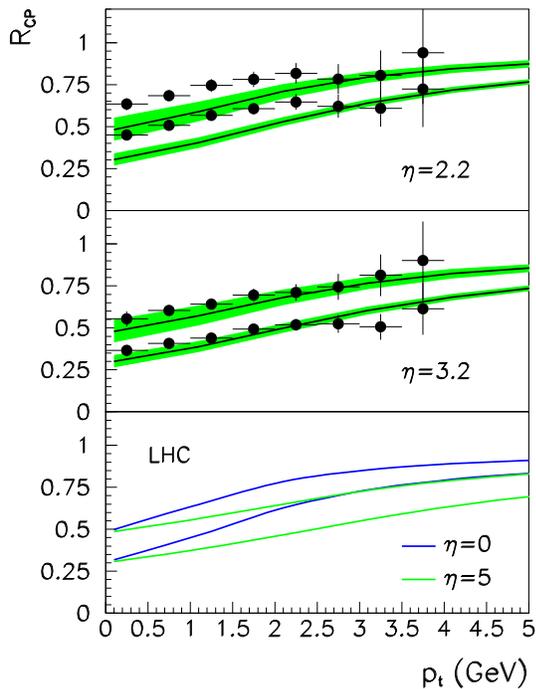}}
\caption{Normalized ratios of central and semi-central to peripheral dAu 
collisions measured by BRAHMS \cite{Arsene:2004ux} compared to results from
Eq.~(\ref{eqratpt}). The bands represent the uncertainty in the determination
of $N_{\rm coll}$~\cite{Arsene:2004ux}.
Results for the same centrality classes at the LHC are given in the lower 
panel.}
\label{figbrahms}
\end{figure}

\noindent 4. In the current debate of RHIC data on the suppressed high-$p_t$ 
hadroproduction in nuclear collisions, the relevance of nuclear 
shadowing has been discussed repeatedly~\cite{Albacete:2003iq,others}. 
It is clear by now~\cite{Gyulassy:2004zy}
that the $A$-dependence of $p_t$-differential hadroproduction  
in nucleus-nucleus collisions and in deuteron-nucleus collisions
at mid-rapidity both involve additional nuclear effects which are
at least as significant as nuclear shadowing. On the other hand,
arguments have been put forward~\cite{Albacete:2003iq,others}
that in d+Au collisions at forward
rapidity, nuclear shadowing may be the dominant effect. Motivated by 
the phenomenological success of the scaling ansatz (\ref{eqmult}), 
we now test to what extent the centrality dependence of $p_t$-differential 
hadron spectra in d+Au emerges naturally from the geometric scaling
found in $\gamma^*A$. We start from the model (\ref{eqfact}), using
for the $k_t$-differential nuclear gluon distribution the scaling function 
$\Phi(\tau)$ at virtuality $\Phi(\tau) \simeq \phi(k=Q/2)$, such that
$\tau=k^2/4\bar Q_{\rm sat,A}^2$, $\bar Q_{\rm sat,A}^2= N_c\qsa^2/C_F$. 
This approximation can be motivated
in a momentum space representation~\cite{probmom} of (\ref{eq1}). 
The parton distribution in the
deuteron is taken to fall off sufficiently quickly, 
$\sim 1/k_t^n$, $n\gg 1$, so that we can write
\begin{equation}
   \frac{\frac{dN^{\rm dAu}_{c_1}}{N_{\rm coll_1}d\eta d^2p_t}}
   {\frac{dN^{\rm dAu}_{c_2}  }{N_{\rm coll_2}d\eta d^2p_t}}
   \approx
   \frac{N_{\rm coll_2}\phi_A(p_t/Q_{\rm sat_1})}
    {N_{\rm coll_1}\phi_A(p_t/Q_{\rm sat_2})}
   \approx
   \frac{N_{\rm coll_2}\Phi(\tau_1)}{N_{\rm coll_1}\Phi(\tau_2)}\, .
   \label{eqratpt}
\end{equation}
We see the use of this rough pocket formula mainly in emphasizing the 
plausible claim that the suppression of d+Au at forward rapidity traces 
directly the suppression of nuclear parton distributions at small $x$. 
For the comparison in Fig.~\ref{figbrahms} to data~\cite{Arsene:2004ux} 
on the normalized yields of central and semi-central over peripheral 
dAu collisions, we use the number of collisions $N_{\rm coll}$ in different 
centrality bins~\cite{Arsene:2004ux} $13.6\pm0.3$, $7.9\pm0.4$ and 
$3.3\pm0.4$. Only the two most forward rapidities $\eta=2.2$ and 3.2 
are compared. We find that Eq.~(\ref{eqratpt}) captures main features 
of the recent data by BRAHMS~\cite{Arsene:2004ux} but it
shows a weaker rapidity dependence.  A more quantitative discussion 
is certainly beyond the accuracy of (\ref{eqratpt}). The only conclusion 
from this exercise is that the more differential analysis of (\ref{eqfact}) 
is not inconsistent with data in d+Au.

\noindent 5.
We now comment on the differences with other approaches. The geometric
scaling in $\gamma^*A$ data has been studied in~\cite{Freund:2002ux},
where a growth of $\qsa^2 \propto A^{\alpha}$, $\alpha \le 1/3$
has been found. This disagreement with our finding could have several
origins. First, $0.01<x<0.1$ was allowed in~ \cite{Freund:2002ux}; 
however, in this antishadowing region, we find
no scaling in the data, as expected.  
Moreover, ~\cite{Freund:2002ux} does not modify the variable $\bar x$ for
small-$Q^2$ as done in this work and in the GBW model.  Second,
~\cite{Freund:2002ux} uses $R_A\propto A^{1/3}$ which leads to differences
in particular for small $A$ -- we find a much worse fit in terms of 
$\chi^2$ for such an ansatz. In ~\cite{Freund:2002ux} this dissagreement is
improved by introducing a free parameter $\gamma$ in the $A$-dependent
normalization of the nuclear $F_2^A$ data, $A^{-\gamma-1}$. In our case,
however, this normalization is fixed by a dimensionful quantity given by
the scaling condition~(\ref{eqnormal}).

The multiplicities in Au+Au collisions at RHIC have been
studied~\cite{kln} on the basis of Eq.~(\ref{eqfact}) 
assuming $\qsa^2\propto A^{1/3}$. These authors are lead to
an expression which is Eq.~(\ref{eqmult}) with $\delta = 1$
times an additional factor $\ln(\sqrt{s}^\lambda N_{\rm part}^{1/3})$
argued to come from scaling violations. We note that for the accessible 
range of $A$, $A^{4/9} \sim A^{1/3}\, \ln(A^{1/3})$ -- this is the 
reason why both approaches provide a fair description of the data at RHIC. 
However, the energy dependence in the logarithmic prefactor introduced 
in ~\cite{kln} implies a flatter centrality dependence with increasing 
energy -- this difference to Eq.~(\ref{eqmult}) becomes sizable at
higher $\sqrt{s}$. 

Finally, the connection between the small $x$- and $A$-dependence of 
parton distribution functions, and the suppression of normalized 
yields in d+Au collisions~\cite{Arsene:2004ux} at forward rapidity has 
been discussed in several recent 
works~\cite{Albacete:2003iq,others}.
Eq.~(\ref{eqratpt}) contributes to
this discussion by illustrating to what extent the suppression of 
high-$p_t$ particles in d+Au at RHIC can be accounted for by the 
shadowing in $\gamma^*A$ collisions ($\Phi = \sigma^{\gamma^*A}$).

%

\noindent 6. Here, we have discussed to what extent data for
different collision systems in $\gamma^*A$, $dA$ and $AA$ can be
related through geometric scaling. Our study does not
exclude the possibility that geometric scaling in $\gamma^*p$
and $\gamma^*A$ is a numerical coincidence without any dynamical origin. 
However, geometric scaling also emerges naturally
in non-linear small-$x$ QCD evolution equations~\cite{bkeq,scaling},
which allow to absorb the entire dependence of small-$x$ parton distributions
on energy and geometry into a single dimensionfull quantity, $\qs$.
The data discussed here are currently considered~\cite{Gyulassy:2004zy} 
to provide the main support for such non-linear saturation effects. 
In fact, the scaling function $\Phi$ in~(\ref{eqscalf}) resembles
the asymptotic solution of the Balitsky-Kovchegov (BK) equation: 
it behaves as $\ln(k/\qs)$ [$(\qs^2/k^2)^b$] for small [large] 
$k$~\cite{bkscal,Albacete:2003iq}.
Given that these non-linear evolution equations hold in a novel 
high-density regime of QCD which may become experimentally accessible,
it is of obvious interest to ask whether the connection between 
geometric scaling in the theory and in the data can be
made more quantitative. On the theoretical side, this requires at least 
the study of the impact parameter dependence~\cite{bdeprc} of small-$x$ 
evolution, and the control of higher order effects. 
In particular, running 
coupling effects are known qualitatively to decrease the energy
dependence~\cite{lambdarc} and the $A$-dependence~\cite{adeprc} 
of the saturation scale in comparison to the BK equation at fixed 
coupling (e.g., Refs.~\cite{adeprc} favor a weak 
$A$-dependence $\qsa^2\propto A^{\alpha}$, $\alpha \ll 1/3$ but 
do not address the impact parameter dependence which is
expected to increase the $A$-dependence). While 
an $A$-dependence of $\qsa^2\propto A^{1/3}$ is often assumed~\cite{mv}, 
much stronger ones (such as $\alpha \simeq{2/3}$~\cite{twothirds})
have also been proposed. The present work has analyzed to what extent
data constrain these energy- and $A$-dependences. These constraints have
to be met by non-linear small-$x$ evolution or by any other model which
aims at providing the common dynamical origin for geometric scaling in 
different nuclear collisions.

We thank R.~Baier, M.~Braun, A.~Capella, D.~Kharzeev, A.~Kovner,
L.~McLerran, G.~Roland, 
K.~Rummukainen and H.~Weigert for discussions.

\end{document}